# Globular structure of the hypermineralized tissue in human femoral neck

Qiong Wang[1,2], Tengteng Tang[3], David Cooper[4], Felipe Eltit[5,6], Peter Fratzl[3], Pierre Guy[2,7] and Rizhi Wang*[1,2,8]

[1] Department of Materials Engineering, University of British Columbia, Vancouver, BC, Canada.
[2] Centre for Hip Health and Mobility, Vancouver, BC, Canada.
[3] Department of Biomaterials, Max Planck Institute of Colloids and Interfaces, Potsdam, Germany.
[4] Department of Anatomy, Physiology and Pharmacology, University of Saskatchewan, Saskatoon, Canada
[5] Vancouver Prostate Centre, Vancouver, Canada.
[6] Department of Urologic Sciences, University of British Columbia, Vancouver, BC, Canada.
[7] Department of Orthopaedics, University of British Columbia, Vancouver, BC, Canada
[8] School of Biomedical Engineering, University of British Columbia, Vancouver, BC, Canada

*Corresponding address:
Dr. Rizhi Wang
Department of Materials Engineering
University of British Columbia
309-6350 Stores Road
Vancouver, BC V6T 1Z4
Canada
Email: rzwang@mail.ubc.ca
Phone: +1-604-822-9752

**Highlights**

- We report the ultrastructure of the hypermineralized tissue in human femoral neck from macro- to nano- scale.
- SR micro-CT found larger but less densely distributed cellular lacunae in hypermineralized tissue than in lamellar bone.
- The hypermineralized tissue is composed of apatite globules with sizes varying from submicron to a few microns.
- Apatite mineral are better crystalline at the globular boundaries than inside the globules.
- FIB-SEM 3D shows nanosized channels inside the globules.

# Abstract

Bone becomes more fragile with ageing. Among many structural changes, a thin layer of highly mineralized and brittle tissue covers part of the external surface of the thin femoral neck cortex in older people and has been proposed to increase hip fragility. However, there have been very limited reports on this hypermineralized tissue in the femoral neck, especially on its ultrastructure. Such information is critical to understanding both the mineralization process and its contributions to hip fracture. Here, we use multiple advanced techniques to characterize the ultrastructure of the hypermineralized tissue in the neck across various length scales. Synchrotron radiation micro-CT found larger but less densely distributed cellular lacunae in hypermineralized tissue than in lamellar bone. When examined under FIB-SEM, the hypermineralized tissue was mainly composed of mineral globules with sizes varying from submicron to a few microns. Nano-sized channels were present within the mineral globules and oriented with the surrounding organic matrix. Transmission electron microscopy showed the apatite inside globules were poorly crystalline, while those at the boundaries between the globules had well-defined lattice structure with crystallinity similar to the apatite mineral in lamellar bone. No preferred mineral orientation was observed both inside each globule and at the boundaries. Collectively, we conclude based on these new observations that the hypermineralized tissue is non-lamellar and has less organized mineral, which may contribute to the high brittleness of the tissue.

# 1. Introduction

Hip fractures are the most common musculoskeletal injuries in the aging population resulting in significant morbidity, mortality, and costs (Braithwaite et al., 2003; Johnell and Kanis, 2004). The global prevalence is estimated to rise to 2.6 million by 2025 (Cooper et al., 1992). Among all types of hip fracture, femoral neck fractures are a particular concern since they usually have a higher risk of complications, including a 20-25% mortality rate for women and 30-35% for men during the first year after fracture (Michelson et al., 1995).

Biomechanically, increased bone fragility at the material level as a result of ageing is the intrinsic reason for a hip to fracture upon an external impact. The human femoral neck consists of a trabecular core enclosed in a cortical shell. Its fragility or fracture resistance depends on bone quality, which is governed by multiple factors including total bone mass, three dimensional structure of both the cortical bone and the trabecular bone, degree of mineralization, as well as degree of cross-linking in the collagen network (Burr, 2019; Crabtree et al., 2001; Cummings et al., 2002; Currey et al., 1996; Fratzl-Zelman et al., 2009; Greenspan et al., 2009; Mayhew et al., 2005; Milovanovic et al., 2012; Reeve and Loveridge, 2014; Wang et al., 2002). Although the importance of bone quality has been well-recognized, current strategies for assessing fracture risk all assume that bone's structure in the femoral neck is uniformly lamellar and the degree of collagen mineralization (~70 wt%) in bone is closely regulated through bone remodeling. Histological studies, although limited, suggest that these assumptions are not true locally. Early studies found hypermineralized tissue (HMT) overlying the cortical bone of the femoral neck in seniors (Boyce and Bloebaum, 1993) and postmenopausal women (Vajda and Bloebaum, 1999). In a later study on the mid-femoral neck, the authors reported that 20% - 70% of the surface is covered by HMT (Allen and Burr, 2005). Recently, we found the mineral content of HMT in human femoral necks of the elderly to be 15% higher than that of lamellar bone, and its fracture toughness to be only one-tenth of that in normal bone (Tang et al., 2019). The presence of this brittle HMT on the surface area of femoral neck would increase hip fragility as it may facilitate crack initiation and significantly compromise bone's fracture resistance to impact.

Despite its potential impact on hip fracture risk, the nature of HMT in the femoral neck is poorly understood. Previous studies described it as calcified fibrocartilage or calcified periosteum (Allen and Burr, 2005; Shea et al., 2001). Our recent study confirmed the presence of collagen type II, the main collagenous extracellular matrix constituting cartilage (Tang et al., 2019). Synchrotron small angle X-ray scattering and wide angle X-ray scattering (SAXS/WAXS) found that apatite in HMT in the femoral neck were smaller and less organized (Tang et al., 2019). The mineral seemed to form micron-sized globular agglomerates under a scanning electron microscope. To the best of our knowledge, no further microscopy studies have been reported on these mineral globules. Such information is critical to understand both the mineralization process of HMT and its contributions to hip fracture.

In this study, we characterized the ultrastructure of HMT in the femoral neck of the elderly at various length scales by using multiple technologies. Specifically, synchrotron radiation computed microtomography (SR micro-CT) was employed to compare the size and density of cellular-level lacunae in HMT versus adjacent cortical bone over submillimeter range. Focused ion beam – scanning electron microscopy (FIB-SEM) tomography was used to study the mineral distribution in three dimensions (3D) up to tens of microns. Transmission electron microscopy (TEM) was used to analyze the crystal structure and morphology of mineral at nanoscale.

## 2. Materials and Methods
### 2.1. Samples preparation

Two proximal femurs were included in this study following approval by the Clinical Research Ethics Review Board at the University of British Columbia. Sample A was a right femoral neck retrieved from an 84-year-old female due to hip fracture. Sample B was a left cadaver femoral neck from a 73- year-old male donor, obtained from LifeLegacy Foundation without reported metabolic bone tissue conditions. These two samples were freshly frozen at -20 °C until preparation.

Transverse sections with respect to the femoral neck's long axis were cut at the mid-neck (Fig. 1a, b) by using a water-cooled diamond saw (IsoMet 4000, Buehler, Lake Bluff, IL, USA). The bone section was immersed in 70% ethanol for 24 hr, followed by a sequence of graded ethanol (80%, 90% and 100%) for dehydration (24 hr in each concentration) and finally dried in air. The dehydrated bone sections were then embedded in epoxy resin (Epothin 2, Buehler). The transverse surface was ground with a series of carbide grinding papers and polished down to a finish of 0.05 μm silica suspension. The polished surface was carbon-coated and imaged in backscattered electron (BSE) mode in a scanning electron microscope (SEM, Quanta 650, FEI, Thermo Fisher Scientific, Hillsboro, Oregon, USA) to identify HMT (Fig. 1c).

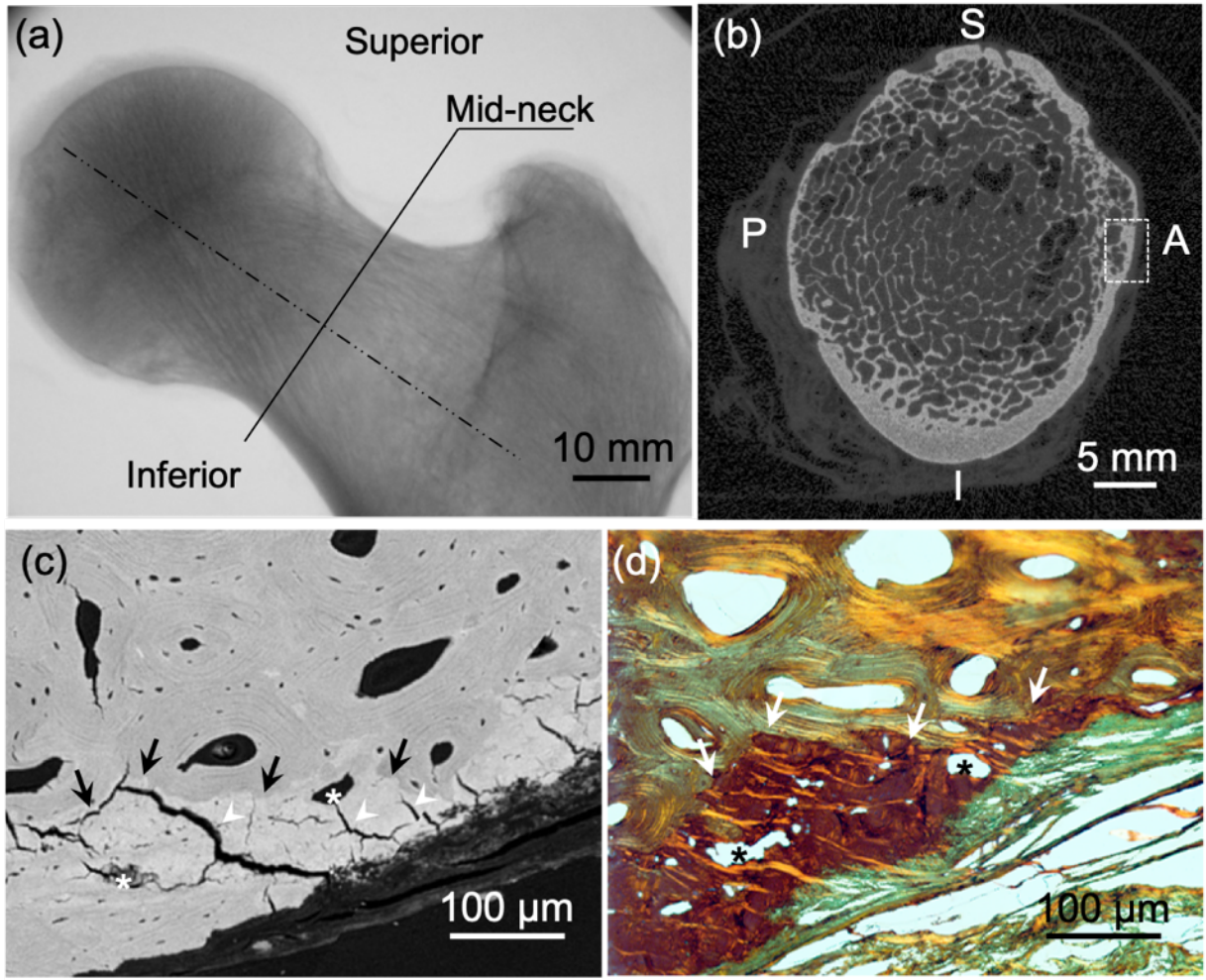

*Figure 1.* Location and histology of the hypermineralized tissue (HMT) of a human femoral neck (73-year-old male). (a) Fluoroscopy image of the proximal femur. (b) HR-pQCT image showing a transverse section of the mid-neck plane showed in (a), S= superior, I = inferior, A= anterior, P = posterior. (c) Backscattered electron (BSE) microscopic image of the lamellar bone and HMT of a region enclosed in the rectangular frame in (b); HMT (delineated by arrows) has brighter appearance and lacks lamellar structure, multiple cracks (arrowheads) and voids (asterisk) are clearly seen. (d) Histological section (polarized light) of the area in (c) stained with picrosirius red; HMT (brown area) is delimited by arrows and voids are highlighted with asterisk.

**2.2. Histology**

In order to observe and confirm the microscopic structure of bone, HMT and the surrounding soft tissues, we prepared samples for histological observations. One section cut from the anterior of sample B (73-year-old male) was fixed in 10% buffered formalin for 24 hr, rinsed in phosphate-buffered saline (PBS) three times for 1 hr each, and demineralized in 10% formic acid for 5 days. The specimens were then dehydrated in an ethanol series (70%, 80%, 90%, 95%, 100%), cleared in xylene substitute, and paraffin embedded. Serial sections of 5 microns thickness were cut with a microtome. The microscopic slides were stained with picrosirius red and Massons' trichrome, to observe distribution of fibrillar collagen under polarized light using an optical microscope (Zeiss Axiophot, Carl Zeiss Microscopy GmbH, Germany).

**2.3. Synchrotron Radiation micro-Computed Tomography (SR micro-CT)**

SR micro-CT was performed on one section cut from the superior-anterior region of sample A (84-year-old female) at the BioMedical Imaging and Therapy (BMIT) facility of the Canadian Light Source synchrotron. The sample was scanned on the BMIT bending magnet beamline using filtered (0.88mm Al + 0.08mm Mo) white beam mode with a mean energy of 26 keV. The sample was mounted 5 cm from the detector to facilitate the development of in-line phase contrast. Projection images (n=3000) were collected over an angular range of 180° with an exposure time of 150 ms each. The CCD camera employed was coupled to a microscopic beam monitor with a 5 × objective lens resulting in a maximum field of view of 3.7 mm x 3.1 mm and a pixel size of 1.44 µm. The 3D dataset was reconstructed using the UFO-Kit software (Vogelgesang et al., 2012) with phase retrieval and then converted from 32- to 8-bit using ImageJ (https://imagej.nih.gov/ij/).

Using ImageJ, a volume of interest approximately 1.2 mm x 0.6 mm x 0.5 mm was isolated for analysis. Segmentation masks separating the bone from the HMT were created using a 3D median filtered (uniform 3D radius of 4 voxels) copy of this VOI. Separation of the tissues was semi-automatically achieved on every 10th image slice utilizing the 2D wand tool followed by manual adjustments where necessary. The intervening slices were segmented by interpolation and checked for accuracy. Prior to quantitative analysis the external mineralizing border of the HMT segmentation mask was contracted 10 voxels and all other borders were contracted 1 voxel to avoid edge artifacts and partial volume effects. Automatic thresholding using the maximum entropy function in ImageJ was utilized to isolate porous spaces in both tissues from the original (unfiltered) images. 3D object counting and volume assessment was employed on all objects larger than 10 voxels (~30 μm$^3$), below which structures were assumed to be noise. Mean lacunar volume (μm$^3$) was directly assessed and lacunar density (lacunae/mm$^3$) was calculated as the number of lacunae divided by the tissue volume. For bone, the vascular canals were subtracted from the tissue volume prior to density calculation. Due to the limited understanding on the nature of HMT, the possible non-lacunar voids were not excluded in the analysis. 3D rendering was performed with CTVox (Bruker Micro-CT, Kontich, Belgium).

### 2.4. FIB-SEM tomography

In order to visualize and examine the 3D architecture of mineral aggregates at the mineralization front of HMT, we used focused ion beam-scanning electron microscopy in serial surface imaging mode (FIB-SEM) with a Zeiss crossbeam 540 station (Carl Zeiss Microscopy, Konstanz, Germany). A backscattered electron image was taken on the surface of one section from superior-anterior of sample A to identify regions containing mineral aggregates at the mineralization front and the specimen was oriented inside the FIB-SEM chamber so that the viewing direction was aligned approximately along the mineralization front. For the milling process with the FIB, a 30 nA gallium beam at 30 kV acceleration voltage was used to create a coarse rectangular section providing a viewing channel for SEM observation. The exposed surface of this rectangular section was then fine-polished by an ion beam current of 3 nA. Finally, the fine-polished block was serially milled using an ion beam of 100 pA at 30 kV. With each removal of each tissue slice, the freshly exposed surface was imaged at 1.7 kV acceleration voltage and 600 pA using both secondary electron (SE) and energy selective backscattered (EsB) detectors. Each slice thickness was roughly equivalent to the lateral resolution of 2D images (~7 nm). The milling process was combined with SEM imaging in sequence (imaging, then sectioning and reimaging) in an automated procedure to collect thousands of serial images. The obtained serial images were aligned (image registration) with in-house python script in Anaconda (Anaconda Inc. Austin, TX, USA) followed by curtaining effects reduction and noise removal. The resulting stacks of images were then segmented based on their structural features and reconstructed using Amira-Avizo (v 2020.1; Thermo Fisher Scientific and Zuse Institute, Berlin, Germany) and Dragonfly (v 4.1; Object Research System Inc. Montreal, Canada).

### 2.5. Scanning transmission electron microscopy (STEM)

To study the ultrastructure of HMT, we prepared 200-nm-thick specimen by using the focused ion beam (FIB) milling and lift-out technique in a dual beam electron microscope (Helios NanoLab DualBeam 650, FEI). FIB milling provides precise control of the area of interest in the specimen. The carbon-coated specimen (section from superior-anterior location of sample B) was imaged first in the backscattered electron mode to locate the interface of lamellar bone and HMT. A platinum (Pt) layer with approximate dimensions of 20 μm long × 2 μm wide × 2 μm thick was deposited to mark the region of interest and to protect the underlying tissue from beam damage. A FIB beam current of 2.5 nA, 30 kV was first used to mill two trenches on either side of Pt layer. The thin layer of tissue between these two trenches was then milled and polished with a series of decreasing currents. The final tissue section was approximate 20 μm long × 2 μm wide × 10 μm deep and was lifted out and mounted on a semi-grid. Afterwards, an ion acceleration voltage of 15 kV and a beam current of 0.43 nA was used to further mill the tissue to approximately 200 nm thick for STEM analysis. Final polishing was performed with an ion acceleration voltage of 15 kV and a beam current of 50 pA. The ultrastructure of the prepared section was analyzed using a FEI Tecnai Osiris STEM operated at 200 kV. The section was analyzed by bight-field (BF), dark-field (DF), as well as high-angle annular dark field (HAADF) imaging in the STEM. Selected area electron diffraction (SAED) with an aperture of 900 nm in diameter was used to analyze crystal structure and orientation.

3. **Results**
   **3.1. General histology**

On the periosteal surface region of the femoral neck, HMT appears brighter than lamellar bone under SEM-BSE mode, indicating a higher mineral content (Fig. 1c). A lack of lamellar structure is clearly shown in HMT, further distinguishing it from lamellar bone. HMT is mostly covered by soft tissue on the surface. The surface of HMT is highly irregular with large pits extending well into the tissue. Histological sections stained with picrosirius red and observed under polarized light, revealed that high birefringence was clearly seen in the lamellar bone and adjacent soft tissue, but not in HMT (Fig. 1d), indicating the matrix in HMT does not have its fibrillary nature. These results confirm previous observations and suggests that the higher mineral density in HMT is accompanied with a decrease in the fibrillar component of the tissue (Tang et al., 2019).

   **3.2. Synchrotron Radiation micro-Computed Tomography (SR micro-CT)**

With SR micro-CT, bone and the HMT appear distinct due to their differential X-ray contrast (Fig. 2a). The smooth endosteal surface of the bone starkly contrasts with the irregular pitted surface of the HMT (Fig. 2c). The two tissues share a complex and convoluted border with structures of Haversian systems penetrating from the bone into the HMT (Fig. 2d). The central canals of these Haversian-like structures taper to blind ends, suggesting a premature halt to the remodeling process. Cellular-scale lacunae occur in both tissues (Fig. 2b), however those within the HMT are more variable and irregular with an average volume approximately 27 times larger (6888 µm$^3$) than that of the osteocyte lacunae within bone (255 µm$^3$). The volumetric density of cellular lacunae is, conversely, approximately 6 times higher in bone (27590 lacunae/mm$^3$) when compared to the HMT (4919 lacunae/mm$^3$) with the overall percent lacunar porosity in the tissues being 0.70% and 3.4% respectively.

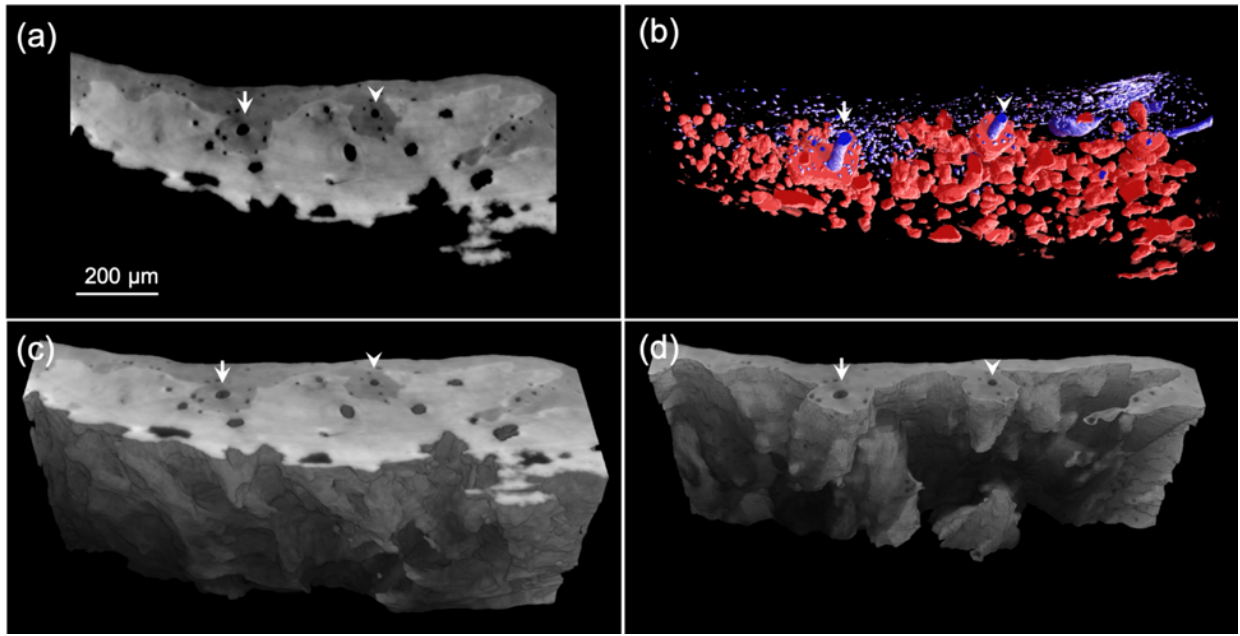

*Figure 2.* (a) Single 2D SR µCT image from the top of the volume of interest analyzed for sample A. Bone appears at the top of the image with the endosteal surface facing upward. Extensions of Haversian systems in bone (grey region)–penetrate into the HMT below (bright region). The central canals of these structures taper off and end blindly. The surface of the HMT (facing the bottom) is highly irregular with deep pits. (b) 3D rendering of isolated cellular-level lacunar porosity in superior view. Osteocyte lacunae and vascular canals within the bone are rendered in blue while lacunae within the HMT are rendered in red. (c) 3D rendering in oblique view rotated 45 degrees downward to reveal the HMT surface. (d) 3D rendering of the irregular bone interface with the HMT (HMT fully transparent) in the same view as (c). Haversian system-like structures protrude out from the bone surface. The arrowhead and arrow indicate two Haversian canals in bone.

### 3.3. Three-dimensional ultrastructure of mineral globules by FIB-SEM tomography

At the mineralization front of HMT (Fig. 3a), aggregates of mineral globules in micron size could be clearly identified in the BSE image (arrowheads). Such mineral was better visualized by 3D volume rendering where mineral globules of various sizes appeared dispersed in the dark matrix (Fig. 3b). Most of the mineral globules tended to merge into bigger clusters, as is shown in Fig. 3c (blue) and the Supplementary Movie S1. The surface of these micron-sized mineral globules appeared irregular and uneven (Fig. 3c, arrow). Additionally, nano-particles in spherical/oval shapes and with smooth surface and dense core were also present (Fig. 3c, arrowhead).

High resolution FIB-SEM data with a voxel size of 7.0 x 7.0 x 8.0 $nm^3$ revealed interesting dark, fine pore features associated with the mineral globules (Fig. 3d insert, arrowheads). These pores with a diameter of ~30 nm primarily appeared in the peripheral region of the mineral globules. 3D surface rendering of these pores showed that they formed channel/tubule-like structures (Fig. 3d, red and Movie S2) with a preferred orientation. Such structural feature can be clearly identified in the transverse view (Fig. 3e) where the dark, long channel structure was visible is a sequence of FIB-SEM images (Fig. 3e, arrowheads) and largely following the direction of the fibres of the underlying matrix (Fig. 3e, arrows).

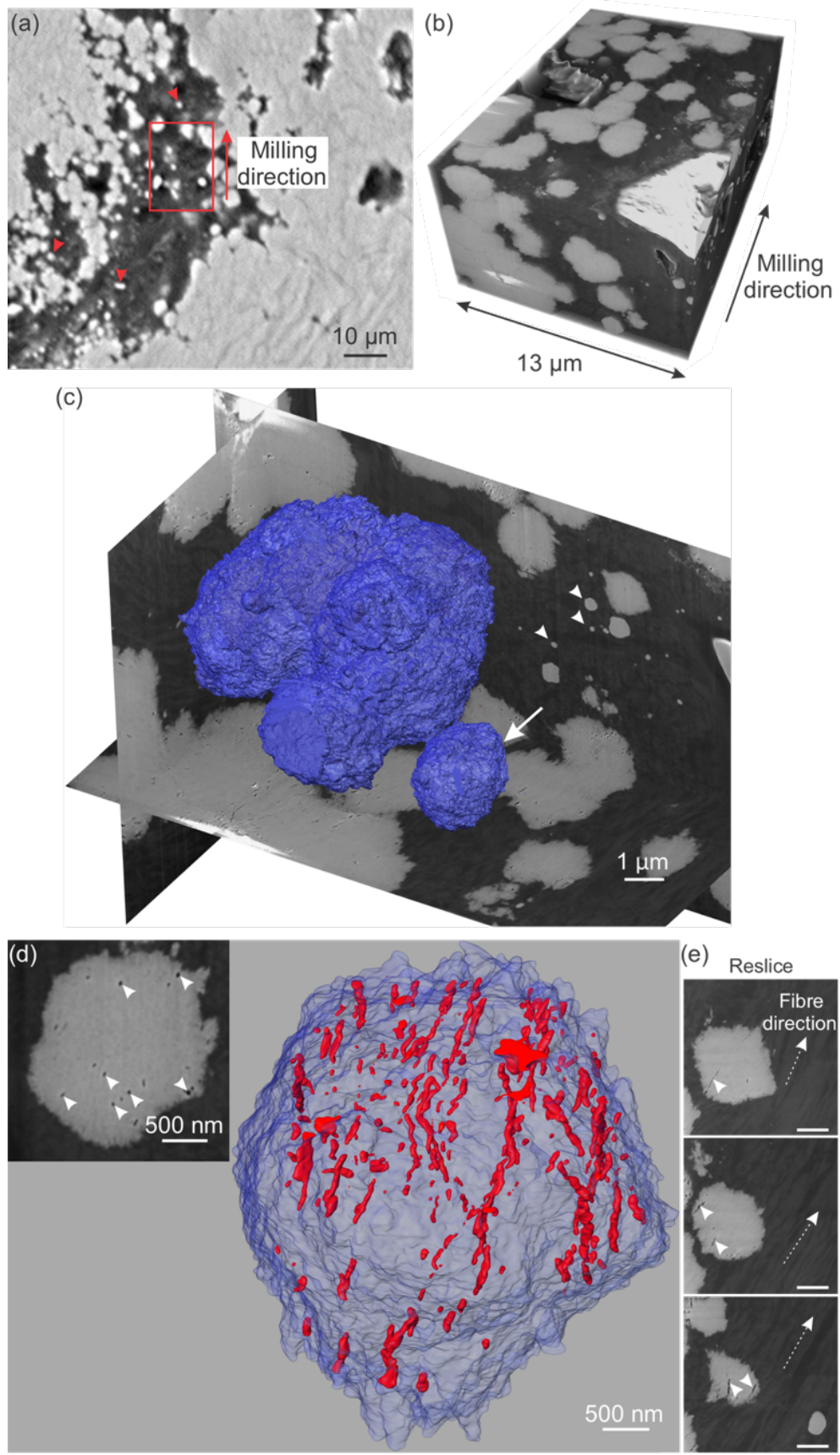

*Figure 3.* Three-dimensional (3D) structure of mineral globules at the mineralization front of HMT. (a) BSE image showing multiple mineral globules (arrowheads) of HMT dispersed in the organic matrix. (b) FIB-SEM 3D volume rendering of (a) (red frame) illustrating numerous mineral aggregates (white). (c) 3D surface rendering corresponding to (b) of the aggregates of mineral globules (blue) intersecting with three orthogonal FIB-SEM image planes demonstrating the rough surface of the mineral and the spherical shape of individual mineral globule (arrow). Nano-sized mineral particles with smooth surfaces probably originating from mineralized matrix vesicles were also observed (arrowheads) (d) FIB-SEM image of a single globule (insert) showing dark, fine pores within the mineral (arrowheads). The 3D reconstruction of the pores (red) and the mineral globule (light blue) delineating the tube/channel-like structures within the mineral. (e) Sequence of transverse views obtained by digital processing (reslice) the full stack of FIB-SEM slices through the same tissue volume represented by the SE image in (d). The dark tube/channel-like structures (arrowheads) are visible and primarily following the direction of the fibres of the non-mineralized organic matrix (arrows). Scale bar = 500 nm.

### 3.4. STEM observation of mineral crystallinity and orientation in HMT

The morphological differences observed between lamellar bone and HMT, led us to analyze the mineral phase of the tissues in depth using STEM. We first located HMT in the periosteal region that appeared brighter than lamellar bone under BSE-SEM (Fig. 4a). On the mineralization front, clusters of mineral globules at the micron scale could be seen dispersed in the soft tissue (Fig. 4a circle). The size of mineral globules increased as they merged into the dense HMT (Fig. 4b). The interface of lamellar bone and HMT was located and prepared for STEM analysis (Fig. 4c).

We observed clear structural differences between HMT and its adjacent lamellar bone through HAADF in STEM (Fig.5a; Fig. S1). In lamellar bone, a banding pattern from the mineralized collagen fibrils could be seen (Fig. 5a, lower right corner). At high magnification, the crystals in the lamellar bone (Fig. 5a, region b) showed well-defined lattice fringes (Fig. 5b, arrowheads), and the electron diffraction pattern showed a preferred orientation on the (002) plane of the apatite crystals. There were two types of mineral morphologies in HMT. The first type existed at the boundaries between the mineral globules (Fig. 5a, region c). Apatite mineral showed well-recognized lattice fringes (Fig. 5c, arrowheads), but without a preferred orientation (Fig. 5c, diffraction pattern). The second type of mineral morphology existed inside each globule (Fig. 5a, region d). Lattice fringes were barely identified (Fig. 5d), and much weak apatite diffraction rings were detected (Fig. 5d, diffraction pattern), suggesting less crystalline nature of the apatite crystals compared with mineral in the lamellar bone.

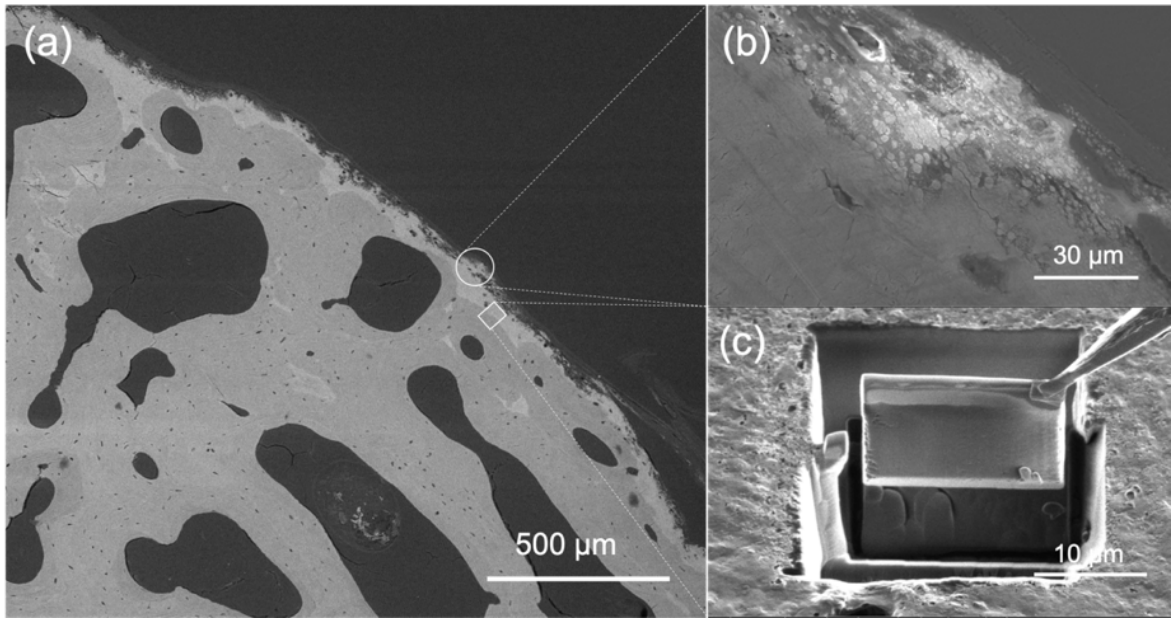

*Figure 4.* TEM sample preparation by FIB cutting and lift out technology. (a) BSE image showing bright HMT on the periosteal region of the cortical bone; (b) High resolution SEM image of the mineralization front of HMT, as circled in (a), clusters of mineral globules are clearly seen; (c) A TEM specimen was prepared and lifted out at the interface between lamellar bone and HMT, indicated in square in (a).

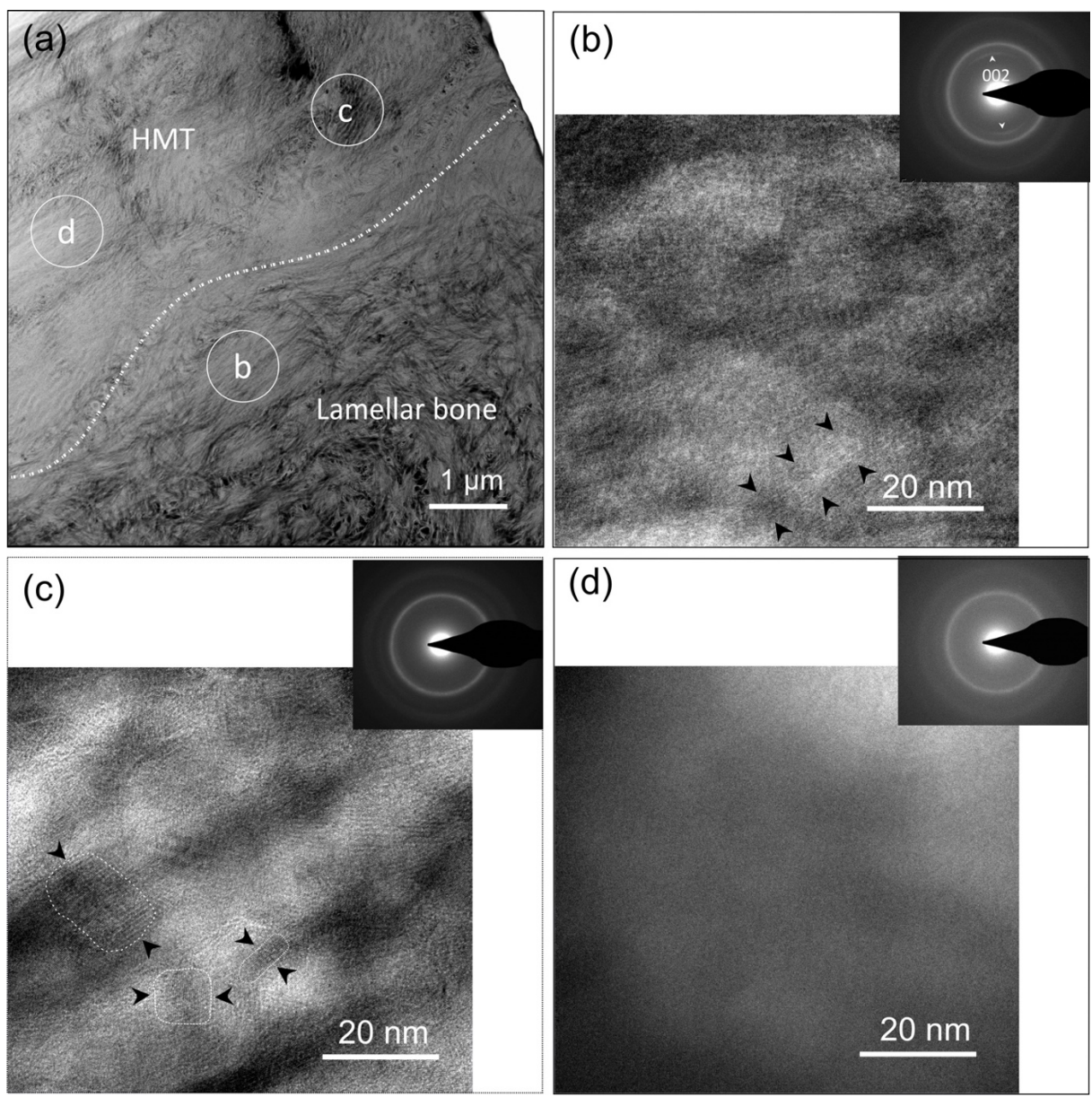

*Figure 5*. TEM analyses of HMT and the adjacent lamellar bone in human femoral neck. (a) STEM Dark Field (HAADF) imaging showing structural differences between lamellar bone (lower right corner) and HMT (upper left corner). The dash line indicates the boundary between lamellar bone and HMT. Three regions of interests (lamellar bone, mineral between globules in HMT, mineral within globules in HMT) are marked in circles. (b) High magnification TEM image and diffraction pattern from selected area (region b) shows recognizable lattice fringes (arrowheads) in lamellar bone. Electron diffraction pattern showed a preferred orientation of the (002) plane. (c) Mineral at the boundary between globules (region c) was crystalline with well-recognized lattice fringes (arrowheads) but random orientation. Individual crystals delimited by a single lattice fringe pattern are contoured by lines. (d) Mineral in the center of globule (region d) showed poor crystalline nature and random orientation based on high magnification TEM image and electron diffraction pattern.

## 4. Discussion

The presence of the hypermineralized tissue around the femoral neck could be a risk factor in hip fracture. There are very limited reports on this tissue, especially on the ultrastructure and nature of mineral within. In this study, our in-depth analyses in this study found that the HMT contained larger but less densely distributed cellular lacunae than lamellar bone. We also observed growing mineral globules with nano-sized channel-like structures in HMT. Apatite crystals are randomly oriented in each globule, with those inside the globule less crystalline than those at the boundaries. These observations provide insights on both the origin and mineralization process of the tissue.

SR-μCT showed cellular lacunae in HMT have great volumetric variations. The average volume of cellular lacunae in HMT is almost 27 times of the lacunae in bone. Since our previous histological study showed the cartilaginous nature of HMT (Tang et al., 2019), we believed these lacunae were from chondrocytes. Even though the lacunae in HMT are larger than the chondrocytes in normal human articular cartilages (Hunziker et al., 2002), the enlargement of hypertrophic chondrocytes is well known in the mineralization of epiphyseal cartilage (Amizuka, 2012; Cooper et al., 2013; Lee et al., 1996). The larger but less densely distributed cellular lacunae in HMT indicated these cells were similar to enlarged chondrocytes. However, all the specimens studied in this report are from seniors and the growth cartilages have been replaced by bone. The origin of this cartilage tissue in the femoral neck of seniors is still unclear. The functions of this HMT are not clear either. Since human femoral neck has significantly less cellular periosteum coverage (only 18.4% surface coverage compared to 59.2% in femoral diaphysis) (Allen and Burr, 2005; Power et al., 2003), periosteal lamellar bone growth would be limited. Therefore, the formation of the hypermineralized tissue on the periosteal surface might be a way to compensate for the bone loss and cortical thinning with ageing or to facilitate growth of femoral neck in order to resist increased stress. Evidences to support this discussion could be found in the SR μCT image (Fig. 2d), where two Haversian systems can be seen penetrating from the bone into the hypermineralized tissue, indicating possible remodeling of the HMT into lamellar bone.

One interesting observation is the small dark pores (~30 nm) discernible within the mineral globules (Fig. 3d, Movie S2) in high resolution FIB-SEM imaging. These pores formed channel/tubule like structure in 3D and largely following the direction of the matrix fibres (Fig. 3e). There have been no reports of this feature in the literature on human bone, and the nature and the potential functions of these pores are unknown at this stage. However, similar features were observed in other normally mineralizing biological tissues, including turkey leg tendon and fish pharyngeal jaw. In turkey leg tendon, a significant number of fine pores with a diameter of ~40 nm were found connected to the canaliculi, comprising an extensive extracellular network through the mineralizing matrix (Zou et al., 2020). The channels formed by these pores are predominantly along the longitudinal direction of collagen fibril bundles and might facilitate the mineralization of collagen fibrils that are far from the tenocytes of the tissue. In a recent study on fish pharyngeal jaw bone, by comparing the mineralized and demineralized specimens, Raguin et al. (Raguin et al., 2020) reported dark channels of approximately 40-50 nm surrounding the mineralized collagen fibrils. These elongated round channels were shown to contain organic material of an unknown nature. With our current FIB-SEM analysis, it is difficult to determine whether these pores are filled with fluid or may contain organic materials such as glycosaminoglycans that inhibit local mineralization. It is interesting to note, however, most of the pores appear in the peripheral region of the individual mineral globule and the channel-like structure clearly correlates with the surrounding oriented organic matrix (Fig. 3e). Such an observation indicates a close relationship between the formation of the pore structure and the surrounding organic tissue. It is plausible that the pore/channel structure could provide space and fluid environment for the transportation of mineral ions and/or mineralization precursors to facilitate nucleation and development of the mineral globules at the mineralization front of HMT.

Another feature of this HMT is the globular aggregation of mineral. The mineral dispersed in the soft tissue at the mineralization front and merged into globules of micrometer size in HMT (Fig. 4b). Morphologically, the mineralization front of HMT in this study looks similar to that of articular calcified cartilage under a scanning electron microscope (Ferguson et al., 2003; Gupta et al., 2005). However, we are not aware of any reports that specifically noted those micron-sized globular features shown in Figs. 3a and 4b. Previous synchrotron small angle X-ray scattering and wide angle X-ray scattering (SAXS/WAXS) found that on average the apatite mineral in HMT was thinner and shorter than that in lamellar bone (Tang et al., 2019). Our STEM results here showed further variations. The mineral inside the micro-globules was poorly crystalline (Fig. 5d). Interestingly, at the boundary of these globules, the mineral became better crystallized but still randomly organized (Fig. 5c). This feature is probably associated with the mineralized matrix vesicles *in vivo*. At the mineralization front, nano-sized mineral spheres with smooth boundary and dense core were observed in the matrix (Fig.3c, arrowheads), which are different from the

micro- globules with pores and rough surfaces (Fig. 3c, arrow and Movie S1). These mineral spheres possibly originated from matrix vesicles, the spherical bodies bounded by a lipid bilayer, which play a key role in the mineralization of cartilage and skeletal tissue (Anderson, 1984; Anderson et al., 2005; Bonucci, 1970; Golub, 2009, n.d.; Hasegawa et al., 2017). One TEM study showed the aggregation of mineral vesicles around chondrocytes at the mineralization front of mice calcified cartilage, indicating a mineralization process regulated by chondrocytes (Bullough and Jagannath, 1983). Similar mineral globules/vesicles were also observed in the calcified epiphyseal cartilage of pig (Bonucci, 1970). In addition, reports showed that the initial mineral formed inside the matrix vesicle was amorphous calcium phosphate (Gay et al., 1978; Wu et al., 1993). In later stage, they continued to develop after the rupture of the membrane to form the needle-shaped mineral crystals (Anderson et al., 2005; Hasegawa et al., 2017). In this study, the poorly crystalline mineral observed inside the globules is probably from the early stage of mineralized vesicles, while the random mineral crystals at the boundary are from the later stage of mineral growth.

The presence of the brittle and highly mineralized tissue at the human femoral neck indicates the limitation of using bone mineral density (BMD) for clinical assessment of hip fracture risk. BMD is typically measured by dual energy X-ray absorptiometry (DXA) for diagnoses of osteoporosis. Patients with large amount of HMT compared to normal lamellar cortical bone at their hips might have increased hip fracture risk, yet it's highly likely that they have normal BMD due to the higher mineral content in HMT. That, in part, might explain why the occurrence of hip fractures is not solely associated with osteoporosis diagnosed by DXA. For instance, a recent study that followed over 7,000 volunteers over 7 years found that only 44 % of women and 21 % of men with non-vertebral osteoporotic fractures are clinically diagnosed as osteoporosis (Schuit et al., 2004). In addition, 50 % of elderly hip fracture cases are not classified as being osteoporotic by densitometry (Kanis et al., 2012); and while some populations have been frequently diagnosed as being osteoporotic by standards, they have relative lower hip fracture incidents (Aspray et al., 1996).

To sum up, our study shows HMT is composed of randomly organized mineral globules and lacks the fibrillar components. Fibrillar collagens and lamellar structure are known to provide a toughening mechanisms to mineralized tissues (Ebacher and Wang, 2009; Launey et al., 2010). The lack of such a fibrillar matrix certainly increases the risk of bone fracture, especially in seniors.

In general, bone mass and density decrease with age. However, this decrease is much more substantial for the cortical shell than the trabecular core (Høiseth et al., 1990). While the relative role of trabecular and cortical bone in hip fracture is a topic for debate, a marked cortical thinning in the superior cortex of the femoral neck has been clearly identified as a structurally weak point (Johannesdottir et al., 2011; Yang et al., 2008; Zebaze et al., 2005). The brittle hypermineralized tissue on the surface of thinning superior cortex of femoral neck could further increase the risk of fracture in femoral neck and is worth further investigation.

5. **Conclusion**

HMT on the periosteal surface of human femoral necks is composed of densely packed mineral globules of submicron to a few microns in size. Apatite mineral is randomly organized; they are poorly crystalline inside the globules as compared to that at the boundaries. Oriented nano-channels exist inside the globules. HMT has larger but less densely distributed lacunae than in lamellar bone. HMT is therefore structurally more isotropic and has higher degree of mineralization than lamellar bone.

**Acknowledgements**

This study was supported by Canadian Institutes of Health Research. Part of the research described in this paper was performed at the Canadian Light Source (CLS), a national research facility of the University of Saskatchewan, which is supported by the Canada Foundation for Innovation (CFI), the Natural Sciences and Engineering Research Council (NSERC), the National Research Council (NRC), the Canadian Institutes of Health Research (CIHR), the Government of Saskatchewan, and the University of Saskatchewan.

We wish to thank Dr. Marc Grynpas at Lunenfeld-Tanenbaum Research Institute for insight discussions on bone histology. The authors are grateful to Dr. Clemens Schmitt for technical support with FIB-SEM imaging, Heike Runge for sample carbon coating and Birgit Schonert for sample polishing (all at the Max Planck Institute of Colloids and Interfaces, Golm, Germany).The authors are particularly grateful for the assistance of the staff of the BioMedical Imaging and Therapy facility at Canadian Light Source (CLS) for their assistance and support as well as members of Dr. Cooper's laboratory who were involved with the collection and processing of the SR uCT data including Kushagra Parolia, Gavin King and Dr. Erika Sales. The authors would like to thank 4D Lab at Simon Fraser University for the technique support for the STEM study, and Dr. Danmei Liu at the Centre for Hip Health and Mobility for sample handling.

# References

Allen, M.R., Burr, D.B., 2005. Human femoral neck has less cellular periosteum, and more mineralized periosteum, than femoral diaphyseal bone. Bone 36, 311–316. doi:10.1016/j.bone.2004.10.013

Amizuka, N., 2012. Histology of epiphyseal cartilage calcification and endochondral ossification. Front. Biosci. E4, 526. doi:10.2741/e526

Anderson, H.C., 1984. Mineralization by Matrix Vesicles. Scan. Electron Microsc. v, 953–964.

Anderson, H.C., Garimella, R., Tague, S.E., 2005. The role of matrix vesicles in growth plate development and biomineralization. Front. Biosci. doi:10.2741/1576

Aspray, T.J., Prentice, A., Cole, T.J., Sawo, Y., Reeve, J., Francis, R.M., 1996. Low bone mineral content is common but osteoporotic fractures are rare in elderly rural gambian women. J. Bone Miner. Res. 11, 1019–1025. doi:10.1002/jbmr.5650110720

Bonucci, E., 1970. Fine structure and histochemistry of "calcifying globules" in epiphyseal cartilage. Zeitschrift für Zellforsch. und Mikroskopische Anat. 103, 192–217. doi:10.1007/BF00337312

Boyce, T.M.M., Bloebaum, R.D.D., 1993. Cortical aging differences and fracture implications for the human femoral neck. Bone 14, 769–778. doi:10.1016/8756-3282(93)90209-S

Braithwaite, R.S., Col, N.F., Wong, J.B., 2003. Estimating hip fracture morbidity, mortality and costs. J. Am. Geriatr. Soc. 51, 364–370. doi:10.1046/j.1532-5415.2003.51110.x

Bullough, P., Jagannath, A., 1983. The morphology of the calcification front in articular cartilage. Its significance in joint function. J. Bone Joint Surg. Br. 65-B, 72–78. doi:10.1302/0301-620X.65B1.6337169

Burr, D.B., 2019. Changes in bone matrix properties with aging. Bone 120, 85–93. doi:10.1016/j.bone.2018.10.010

Cooper, C., Campion, G., Melton, L.J., 1992. Hip fractures in the elderly: A world-wide projection. Osteoporos. Int. 2, 285–289. doi:10.1007/BF01623184

Cooper, K.L., Oh, S., Sung, Y., Dasari, R.R., Kirschner, M.W., Tabin, C.J., 2013. Multiple phases of chondrocyte enlargement underlie differences in skeletal proportions. Nature 495, 375–378. doi:10.1038/nature11940

Crabtree, N., Loveridge, N., Parker, M., Rushton, N., Power, J., Bell, K.L., Beck, T.J., Reeve, J., 2001. Intracapsular hip fracture and the region-specific loss of cortical bone: analysis by peripheral quantitative computed tomography. J. Bone Miner. Res. 16, 1318–1328.

doi:10.1359/jbmr.2001.16.7.1318

Cummings, S.R., Bates, D., Black, D.M., 2002. Clinical use of bone densitometry: scientific review. JAMA 288, 1889–97.

Currey, J.D., Brear, K., Zioupos, P., 1996. The effects of ageing and changes in mineral content in degrading the toughness of human femora. J. Biomech. 29, 257–260. doi:10.1016/0021-9290(95)00048-8

Ebacher, V., Wang, R., 2009. A unique microcracking process associated with the inelastic deformation of Haversian bone. Adv. Funct. Mater. 19, 57–66. doi:10.1002/adfm.200801234

Ferguson, V.L., Bushby, A.J., Boyde, A., 2003. Nanomechanical properties and mineral concentration in articular calcified cartilage and subchondral bone. J. Anat. 203, 191–202. doi:10.1046/j.1469-7580.2003.00193.x

Fratzl-Zelman, N., Roschger, P., Gourrier, A., Weber, M., Misof, B.M., Loveridge, N., Reeve, J., Klaushofer, K., Fratzl, P., 2009. Combination of nanoindentation and quantitative backscattered electron imaging revealed altered bone material properties associated with femoral neck fragility. Calcif. Tissue Int. 85, 335–43. doi:10.1007/s00223-009-9289-8

Gay, C. V., Schraer, H., Hargest, T.E., 1978. Ultrastructure of matrix vesicles and mineral in unfixed embryonic bone. Metab. Bone Dis. Relat. Res. 1, 105–108. doi:10.1016/0221-8747(78)90045-0

Golub, E.E., 2009. Role of matrix vesicles in biomineralization. Biochim. Biophys. Acta - Gen. Subj. doi:10.1016/j.bbagen.2009.09.006

Golub, E.E., n.d. Role of Matrix Vesicles in Biomineralization. doi:10.1016/j.bbagen.2009.09.006

Greenspan, S.L., Myers, E.R., Maitland, L.A., Kido, T.H., Krasnow, M.B., Hayes, W.C., 2009. Trochanteric bone mineral density is associated with type of hip fracture in the elderly. J. Bone Miner. Res. 9, 1889–1894. doi:10.1002/jbmr.5650091208

Gupta, H.S., Schratter, S., Tesch, W., Roschger, P., Berzlanovich, A., Schoeberl, T., Klaushofer, K., Fratzl, P., 2005. Two different correlations between nanoindentation modulus and mineral content in the bone–cartilage interface. J. Struct. Biol. 149, 138–148. doi:10.1016/j.jsb.2004.10.010

Hasegawa, T., Yamamoto, T., Tsuchiya, E., Hongo, H., Tsuboi, K., Kudo, A., Abe, M., Yoshida, T., Nagai, T., Khadiza, N., Yokoyama, A., Oda, K., Ozawa, H., de Freitas, P.H.L., Li, M.,

Amizuka, N., 2017. Ultrastructural and biochemical aspects of matrix vesicle-mediated mineralization. Jpn. Dent. Sci. Rev. doi:10.1016/j.jdsr.2016.09.002

Høiseth, A., Alho, A., Husby, T., 1990. Femoral cortical/cancellous bone related to age. Acta radiol. 31, 626–627. doi:10.1177/028418519003100619

Hunziker, E.B., Quinn, T.M., Häuselmann, H.J., Hä, H.-J., Mü, M.E., 2002. Quantitative structural organization of normal adult human articular cartilage. Osteoarthr. Cartil. 10, 564–572. doi:10.1053/joca.2002.0814

Johannesdottir, F., Poole, K.E.S., Reeve, J., Siggeirsdottir, K., Aspelund, T., Mogensen, B., Jonsson, B.Y., Sigurdsson, S., Harris, T.B., Gudnason, V.G., Sigurdsson, G., 2011. Distribution of cortical bone in the femoral neck and hip fracture: A prospective case-control analysis of 143 incident hip fractures; the AGES-REYKJAVIK Study. Bone 48, 1268–1276. doi:10.1016/j.bone.2011.03.776

Johnell, O., Kanis, J.A., 2004. An estimate of the worldwide prevalence, mortality and disability associated with hip fracture. Osteoporos. Int. 15, 897–902. doi:10.1007/s00198-004-1627-0

Kanis, J.A., Odén, A., McCloskey, E. V., Johansson, H., Wahl, D.A., Cooper, C., 2012. A systematic review of hip fracture incidence and probability of fracture worldwide. Osteoporos. Int. 23, 2239–2256. doi:10.1007/s00198-012-1964-3

Launey, M.E., Buehler, M.J., Ritchie, R.O., 2010. On the Mechanistic Origins of Toughness in Bone. Annu. Rev. Mater. Res. 40, 25–53. doi:10.1146/annurev-matsci-070909-104427

Lee, K.M., Fung, K.P., Leung, P.C., Leung, K.S., 1996. Identification and characterization of various differentiate growth plate chondrocytes from porcine by countercurrent centrifugal elutriation. J. Cell. Biochem. 60, 508–520. doi:10.1002/(SICI)1097-4644(19960315)60:4<508::AID-JCB7>3.0.CO;2-W

Mayhew, P.M., Thomas, C.D., Clement, J.G., Loveridge, N., Beck, T.J., Bonfield, W., Burgoyne, C.J., Reeve, J., 2005. Relation between age, femoral neck cortical stability, and hip fracture risk. Lancet 366, 129–135. doi:10.1016/S0140-6736(05)66870-5

Michelson, J.D., Myers, A., Jinnah, R., Cox, Q., Van Natta, M., 1995. Epidemiology of hip fractures among the elderly: Risk factors for fracture type. Clin. Orthop. Relat. Res. 129–135.

Milovanovic, P., Potocnik, J., Djonic, D., Nikolic, S., Zivkovic, V., Djuric, M., Rakocevic, Z., 2012. Age-related deterioration in trabecular bone mechanical properties at material level: Nanoindentation study of the femoral neck in women by using AFM. Exp. Gerontol. 47,

154–159. doi:10.1016/j.exger.2011.11.011

Power, J., Loveridge, N., Rushton, N., Parker, M., Reeve, J., 2003. Evidence for bone formation on the external "periosteal" surface of the femoral neck: a comparison of intracapsular hip fracture cases and controls. Osteoporos. Int. 14, 141–145. doi:10.1007/s00198-002-1333-8

Raguin, E., Rechav, K., Brumfeld, V., Shahar, R., Weiner, S., 2020. Unique three-dimensional structure of a fish pharyngeal jaw subjected to unusually high mechanical loads. J. Struct. Biol. 211, 107530. doi:10.1016/j.jsb.2020.107530

Reeve, J., Loveridge, N., 2014. The fragile elderly hip: Mechanisms associated with age-related loss of strength and toughness. Bone 61, 138–148. doi:10.1016/j.bone.2013.12.034

Schuit, S.C.E., Van Der Klift, M., Weel, A.E.A.M., De Laet, C.E.D.H., Burger, H., Seeman, E., Hofman, A., Uitterlinden, A.G., Van Leeuwen, J.P.T.M., Pols, H.A.P., 2004. Fracture incidence and association with bone mineral density in elderly men and women: The Rotterdam Study. Bone 34, 195–202. doi:10.1016/j.bone.2003.10.001

Shea, J.E., Vajda, E.G., Bloebaum, R.D., 2001. Evidence of a hypermineralised calcified fibrocartilage on the human femoral neck and lesser trochanter. J. Anat. 198, 153–162. doi:10.1046/j.1469-7580.2001.19820153.x

Tang, T., Wagermaier, W., Schuetz, R., Wang, Q., Eltit, F., Fratzl, P., Wang, R., 2019. Hypermineralization in the femoral neck of the elderly. Acta Biomater. 89, 330–342. doi:10.1016/j.actbio.2019.03.020

Vajda, E.G., Bloebaum, R.D., 1999. Age-related hypermineralization in the female proximal human femur. Anat. Rec. 255, 202–211. doi:10.1002/(SICI)1097-0185(19990601)255:2<202::AID-AR10>3.0.CO;2-0

Vogelgesang, M., Chilingaryan, S., Rolo, T.D.S., Kopmann, A., 2012. UFO: A scalable GPU-based image processing framework for on-line monitoring, in: Proceedings of the 14th IEEE International Conference on High Performance Computing and Communications, HPCC-2012 - 9th IEEE International Conference on Embedded Software and Systems, ICESS-2012. IEEE, pp. 824–829. doi:10.1109/HPCC.2012.116

Wang, X., Shen, X., Li, X., Mauli Agrawal, C., 2002. Age-related changes in the collagen network and toughness of bone. Bone 31, 1–7. doi:10.1016/S8756-3282(01)00697-4

Wu, L.N.Y., Yoshimori, T., Genge, B.R., Sauer, G.R., Kirsch, T., Ishikawa, Y., Wuthier, R.E., 1993. Characterization of the nucleational core complex responsible for mineral induction by growth plate cartilage matrix vesicles. J. Biol. Chem. 268, 25084–25094.

Yang, L., Maric, I., McCloskey, E. V., Eastell, R., 2008. Shape, Structural Properties, and Cortical Stability Along the Femoral Neck: A Study Using Clinical QCT. J. Clin. Densitom. 11, 373–382. doi:10.1016/j.jocd.2008.04.008

Zebaze, R.M.D., Jones, A., Welsh, F., Knackstedt, M., Seeman, E., 2005. Femoral neck shape and the spatial distribution of its mineral mass varies with its size: Clinical and biomechanical implications. Bone 37, 243–252. doi:10.1016/j.bone.2005.03.019

Zou, Z., Tang, T., Macías-Sánchez, E., Sviben, S., Landis, W.J., Bertinetti, L., Fratzl, P., 2020. Three-dimensional structural interrelations between cells, extracellular matrix, and mineral in normally mineralizing avian leg tendon. Proc. Natl. Acad. Sci. 117, 14102–14109. doi:10.1073/pnas.1917932117